\documentclass[aps,12pt,showkeys]{revtex4}
\usepackage[sort&compress]{natbib}
 
\usepackage{xcolor}
\usepackage{graphicx}
\usepackage{amsmath}
\usepackage{amssymb}
\usepackage{epstopdf}
\usepackage{float}
\usepackage{hyperref}
\hypersetup{
    colorlinks,
    citecolor=blue,
    filecolor=blue,
    linkcolor=blue,
    urlcolor=blue}  
\begin{document}

\title{Quadruple Beltrami field structures in electron-positron multi-ion plasma}

\author{Farhat Saleem}
\author{M. Iqbal}
\author{Usman Shazad}
\email{usmangondle@gmail.com}
\affiliation{Department of Physics, University of Engineering and Technology, Lahore
54890, Pakistan}

\begin{abstract}
{A quadruple Beltrami (QB) equilibrium state for a four-component plasma that consists of inertial electrons, positrons, lighter positive ($H^{+}$) ions and heavier negative ions ($O_{2}^{-}$) is derived and investigated. The QB relaxed state is a linear superposition of four distinct single Beltrami fields and provides the possibility of the formation of four self-organized vortices of different length scales. In addition, robust magnetofluid coupling characterizes this non-force-free state. The analysis of the QB state also shows that by adjusting the generalized helicities and densities of plasma species, the formation of multiscale structures as well as the paramagnetic and diamagnetic behavior of the relaxed state can be controlled.}
\end{abstract}

\maketitle

\section{Introduction}\label{S1}

Despite the inherent complexity arising from the interaction between magnetic
fields and flows, the ordered behavior of magnetized plasma has been observed
in nature and laboratory plasma experiments. The emergence of the Beltrami magnetic field ($\mathbf{B}$) serves as a well-known illustration of this phenomena. The Beltrami magnetic field, represented by the equation $\mathbf{\nabla}\times\mathbf{B}=\lambda\mathbf{B}$, where $\lambda$ is a constant that can take on real, complex, or even imaginary values, characterizes the state of equilibrium where the energy associated with the flow can be ignored. However, in general $\lambda$ can also be spatially dependent. This equilibrium state is commonly referred to as a relaxed or self-organized state \cite{Ortolani}.
The Beltrami relaxed state, in its essence, unveils the manifestation of the force-free macroscopic state within the magnetoplasma. This state arises as a consequence of the absence of plasma flows and pressure gradients. When applied to a plasma that is perfectly conducting, the variational principle can be used to derive the Beltrami field or relaxed state, which is defined by a constant $\lambda$--eigenvalue of the curl operator that is also termed a scale parameter. In the context of ideal magnetohydrodynamics (MHD), the derivation of Beltrami fields involves the minimization of magnetic energy, subject to the constraints that the local magnetic helicity, a metric for the structural intricacy of magnetic field lines, remains invariant \cite{Woltjer1958}. 
The single Beltrami field or state is alternatively referred to as Taylor's relaxed state, owing to Taylor's conjecture. According to Taylor's conjecture, realistic plasmas exhibit small resistivity, resulting in the dissipation of magnetic energy at a higher rate compared to magnetic helicity. It is also suggested that the magnetic helicity of the plasma is globally conserved \cite{Taylor1974}.

Later on, it has been demonstrated theoretically that in Hall magnetohydrodynamics (HMHD), it is possible to achieve non-force-free (NFF) relaxed states that are more realistic. These NFF self-organized states are characterized by the combination of two distinct single Taylor states or Beltrami fields with a robust flow. These relaxed states can be derived by the topological constrained minimization (helicity constraints) of magnetofluid energy using the variational principle or by employing the vortex dynamics approach \cite{Steinhauer1997,Mahajan1998,Yoshida1999,Steinhauer1998,Steinhauer2002}. Furthermore, Yoshida and Mahajan \cite{Yoshida2002} have highlighted that a magnetized two-fluid plasma undergoes relaxation when its canonical enstrophy (--curl of the generalized or canonical momentum), which serves as an indicator of dissipation and turbulence, is minimized, but the ideal invariants of the plasma, such as magnetofluid energy, magnetic helicity, and generalized helicity, remain constant. It is of utmost significance to acknowledge that within a magnetized multi-fluid plasma comprising $n$ species, there exist $n+1$ ideal invariants. Furthermore, it has been demonstrated that the constrained minimization of these ideal invariants results in the emergence of multi-Beltrami relaxed states \cite{Mahajan2015}.

Importantly, the aforementioned NFF Beltrami states have extensive applications in the modeling of different plasma phenomena in laboratory and astrophysical plasmas, including plasma confinement \cite{Mahajan2000,Yoshida2001}, heating of solar corona \cite{Mahajan2001}, flow generation in subcoronal regions \cite{Mahajan2002}, explosive and eruptive events in solar atmosphere \cite{Ohsaki2002}, solar arcades \cite{Bhattacharyya2007}, solar coronal loops \cite{Kumar2011}, dynamo and reverse dynamo processes \cite{Mahajan2005,Lingam2015}, non-linear Alfv\'{e}n waves and helicons  \cite{Abdelhamid2016a,Abdelhamid2017} and turbulence \cite{Abdelhamid2016b,Mahajan2020}. Over the past several years, there has been an increasing amount of research dedicated to exploring the multi-Beltrami relaxed state and its interesting implications for various plasma models. These include dense degenerate plasmas \cite{Berezhiani2015,Shatashvili2016,Shatashvili2019}, relativistic hot plasmas \cite{Usman2021,Usman2023a,Usman2023b,Usman2024}, general relativistic plasmas in the vicinity of black holes \cite{Bhattacharjee2018,Asenjo2019,Bhattacharjee2020a,Bhattacharjee2020b}, multi-ion and partially ionized space plasmas \cite{Shafa2022,Faheem2023}, as well as plasmas in a massive photon field \cite{Bhattacharjee2023,Usman2023c}.

The main aim of the present investigation is to explore the possibility of a multi-Beltrami relaxed state in a four-component magnetized plasma comprising electrons, positrons, singly ionized light positive ions ($H^{+}$), and heavy negative ions ($O_{2}^{-}$).
Negative ions can be observed in plasmas due to fundamental mechanisms, including the dissociative or non-dissociative attachment of electrons to neutral species \cite{Vladimirov2003}. On the other hand, positrons can be generated in several astrophysical scenarios, where cosmic-ray nuclei interact with atoms \cite{Adriani2009}, as well as in contemporary laser plasma studies, when matter is subjected to ultra-intense laser pulses \cite{Kourakis2006}. Multispecies plasmas, which consist of negative ion species, are frequently observed in diverse space and astrophysical environments. These include the D-F regions of the ionosphere of the earth \cite{Massey1976}, the cometary comae \cite{Chaizy1991}, and the upper atmosphere of Titan \cite{Coates2007}. In addition, positive and negative-ion plasmas can be produced in low-temperature laboratory plasma experiments \cite{Ichiki2002}, neutral beam sources \cite{Bacal1979}, combustion products and plasma etching \cite{Sheehan1988}, and plasma processing reactors \cite{Gottscho1986}. Most importantly, in recent years, several groups of researchers have also investigated the numerous nonlinear plasma phenomena (e.g., waves, solitons, etc.) in electron-positron plasmas with multi-ions in different plasma settings \cite{Sultana2014,Jannat2015,Jannat2016,Chowdhury2017,Ahmed2018,Khondaker2019,Abdelwahed2020,Douanla2020,Jahan2021,Taibany2021,Heera2021}.

In this study, a quadruple Beltrami (QB) relaxed state is derived and investigated using some model plasma parameters relevant to the earth's ionosphere. The QB state is the linear superposition of four force-free single Beltrami fields. It is characterized by four scale parameters, leading to the creation of four self-organized structures with different length scales.  In the self-organized state, the effects of generalized helicities on the formation of multiscale structures, the number densities of plasma species on diamagnetic and paramagnetic trends, and the inertial effects of electron, positron and ion species on the formation of lower index Beltrami states are also investigated. So far as we are aware, this work is novel because it is the first time that the relaxed state of this particular plasma model is being studied.

This article is organized in the subsequent manner: In Sec. \ref{S2}, we present the theoretical model and its corresponding equations that are employed in the derivation of the equilibrium state.
In the ensuing section, we derive an equation representing the equilibrium state in the form of a QB field, and its analytical structure is described. In Sec. \ref{S4}, the QB state is investigated by focusing on model plasma parameters that are relevant to the earth's ionosphere. The final section has been reserved for the summary and conclusion of the present study.
\section{Model equations}\label{S2}
In this investigation, we consider a quasi-neutral, incompressible, collisionless and magnetized four-component plasma comprised of mobile electrons, positrons, light positive ions and heavy negative ions. The quasi-neutrality condition for this plasma system can be mathematically represented as the equation
\begin{equation}
N_{p}+z_{il}N_{il}-z_{ih}N_{ih}\approx 1,
\end{equation}
where $N_{p}=n_{p}/n_{e}$, $N_{il}=n_{il}/n_{e}$ and $N_{ih}=n_{ih}/n_{e}$,
in which $n_{\alpha }$ ($\alpha =e$ -- electrons, $p$ -- positrons, $il$ --
lighter positive ions and $ih$ -- heavier negative ions) is the number density and $z_{\alpha }$ is the charge number of plasma species. Further, the equations of motion that describe the dynamics of plasma species in a dimensionless form can be succinctly stated as
\begin{equation}
\frac{\partial \mathbf{P}_{\alpha }}{\partial t}=\mathbf{V}_{\alpha }\times 
\mathbf{\Omega }_{\alpha }-\mathbf{\nabla }\kappa _{\alpha },\label{em}
\end{equation}
where $\mathbf{P}_{e}=\mathbf{V}_{e}-\mathbf{A}$, $\mathbf{P}_{p}=\mathbf{V}%
_{e}+\mathbf{A}$, $\mathbf{P}_{il}=\mathbf{V}_{il}+z_{il}M_{il}\mathbf{A}$, $%
\mathbf{P}_{ih}=\mathbf{V}_{ih}-z_{ih}M_{ih}\mathbf{A}$, $\mathbf{\Omega }%
_{\alpha }=\mathbf{\nabla }\times \mathbf{P}_{\alpha }$, $\kappa
_{e}=-\varphi+p_{e}+0.5V_{e}^{2}$, $\kappa _{p}=\varphi
+N_{p}^{-1}p_{p}+0.5V_{p}^{2}$, $\kappa _{il}=z_{il}M_{il}\varphi +M_{il}N_{il}^{-1}p_{il}+0.5V_{il}^{2}$, $%
\kappa _{ih}=-z_{ih}M_{ih}\varphi +M_{ih}N_{ih}^{-1}p_{ih}+0.5V_{ih}^{2}$, $M_{il}=m_{e}/m_{il}$ and $M_{ih}=m_{e}/m_{ih}$. The Eq. (\ref{em}) includes the variables $\mathbf{P}_{\alpha}$, $\mathbf{\Omega}_{\alpha}$, $\mathbf{V}_{\alpha}$, $\mathbf{B}$, $m_{\alpha}$ and $p_{\alpha}$, which symbolize generalized momentum, generalized vorticity, velocity, magnetic field, mass and pressure of the plasma species. Additionally, $\varphi$ and $\mathbf{A}$ denote the scalar electric potential and the vector magnetic potential, respectively.
Importantly, to achieve a dimensionless representation of the model equations, we have employed the normalization technique for several variables. Specifically, the length, magnetic field, velocity of plasma species, time, scalar electric potential, vector magnetic potential and thermal pressure have been normalized using following parameters: electron skin depth ($l_{e}=\sqrt{
m_{e}c^{2}/4\pi e^{2}n_{e}}$ , where $c$ is speed of light and $e$ is amount of elementary charge), $B_{0}$--some arbitrary value of ambient magnetic field, Alfv\'{e}n velocity ($c_{A}=B_{0}/\sqrt{4\pi m_{e}n_{e}}$), electron gyroperiod ($l_{e}/c_{A}$), $B_{0}^{2}/4\pi en_{e}$, $\lambda_{e}B_{0}$ and $B_{0}^{2}/4\pi$, respectively. It also is imperative to point out that within the framework of this plasma model, it is assumed that the plasma frequency is significantly higher than the time required for pair annihilation \cite{Tajima1990}. Consequently, the phenomenon of pair annihilation has not been taken into consideration in the scope of this investigation. Furthermore,  by taking the curl of the Eq. (\ref{em}), we derive the vorticity evolution equations that can be represented as
\begin{equation}
\frac{\partial \mathbf{\Omega }_{\alpha }}{\partial t}=\mathbf{\nabla }
\times \left( \mathbf{V}_{\alpha }\times \mathbf{\Omega }_{\alpha }\right).\label{ve}
\end{equation}
Significantly, the gradient terms have vanished and do not contribute to the vortex dynamics of the plasma species. Additionally, Ampere's law is a crucial equation within our mathematical modeling of the plasma system, as it establishes a connection between the independent dynamics of electrons, positrons, light positive ions, and heavy negative ions. The normalized form of Ampere's law, neglecting displacement currents (assuming non-relativistic flow of plasma species) can be represented as the equation
\begin{equation}
\mathbf{\nabla }\times \mathbf{B}=N_{p}\mathbf{V}_{p}+z_{il}N_{il}\mathbf{V}%
_{il}-z_{ih}N_{ih}\mathbf{V}_{ih}-\mathbf{V}_{e}.\label{cd}
\end{equation}
Finally, one can use continuity equations and equations of state for plasma species to close the system of model Eqs. (\ref{em}-\ref{cd}). Prior to delving into the derivations of the relaxed state equations utilizing the aforementioned model equations, it is imperative to emphasize that through the manipulation of Eqs. (\ref{em}-\ref{cd}), one can acquire the constants of motion or ideal invariants for this plasma model. These ideal invariants are the generalized or canonical helicities ($H_{\alpha}$) of plasma species and the magnetofluid energy ($E$), which can be mathematically represented as
\begin{equation}
    H_{\alpha }=\frac{1}{2}\int_{v}\left( \mathbf{P}_{\alpha }\cdot \mathbf{
\Omega }_{\alpha }\right) dv,\label{he}
\end{equation}
\begin{equation}
    E=\frac{1}{2}\int_{v}\left(
V_{e}^{2}+N_{p}V_{p}^{2}+\frac{z_{il}N_{il}}{M_{il}}V_{il}^{2}+\frac{z_{ih}N_{ih}}{M_{ih}}V_{ih}^{2}+B^{2}\right) dv,\label{en}
\end{equation}
where $dv$ is volume element. Furthermore, the ensuing section covers the importance of these ideal invariants in the relaxation process.
\section{QB equilibrium state}\label{S3}
The most basic and fundamental equilibrium solution to vortex dynamics Eq. (\ref{ve}) is provided by the "Beltrami condition," which requires the alignment of generalized vorticities with flows of the respective plasma species ($\mathbf{\Omega }_{\alpha }\parallel \mathbf{V}_{\alpha }$). One well-known and straightforward illustration of such Beltrami conditions is the single Beltrami field or Taylor relaxed state in an ideal MHD flowless plasma, which is represented by the equation $\mathbf{\nabla}\times\mathbf{B}=\lambda\mathbf{B}$. It's also crucial to keep in mind that the Beltrami conditions represent fundamental tenets of physics, wherein the inertial plasma species adhere to the magnetic field lines that have been modified by their flow vorticities, whereas the inertialess plasma species follow the magnetic field lines. So, the steady-state or time-independent solution of vorticity evolution equations yields the following set of Beltrami conditions for plasma species:
\begin{equation}
\mathbf{\nabla }\times \mathbf{V}_{e}-\mathbf{B}=a\mathbf{V}_{e},\label{be}
\end{equation}
\begin{equation}
\mathbf{\nabla }\times \mathbf{V}_{p}+\mathbf{B}=a\mathbf{V}_{p},\label{bp}
\end{equation}
\begin{equation}
\mathbf{\nabla }\times \mathbf{V}_{il}+z_{il}M_{il}\mathbf{B}=b\mathbf{V}
_{il},\label{bl}
\end{equation}
\begin{equation}
\mathbf{\nabla }\times \mathbf{V}_{ih}-z_{ih}M_{ih}\mathbf{B}=c\mathbf{V}
_{ih},\label{bh}
\end{equation}
where $a$, $b$ and $c$ are some real constants dimensionally equal to the inverse of length called Beltrami parameters that are related with the constants of motion of this plasma system i.e., the generalized helicities and magnetofluid energy of plasma species given by Eqs. (\ref{he}-\ref{en}). In addition, these being real-valued constants essentially serve as the measure of generalized helicities. The Beltrami parameters ($a$ for electron and positron species, $b$ and $c$ for lighter and heavier ion species, respectively), being the ratio between the magnitude of the generalized vorticity and the magnitude of the flow velocity, serve to characterize the strength or amplitude of the flow of plasma species. Hence, for the Beltrami parameter having a value larger than one, the flow of plasma species is sub-Alfv\'{e}nic, whereas for the value smaller than one, the flow of plasma species is super-Alfv\'{e}nic. At this point, it is also very important to note that in this study, we have taken into account identical values of generalized helicities for pair species as a result of their equal masses (same value of Beltrami parameter $a$), whereas different values are being considered for ion species ($b$ and $c$). An additional characteristic of the Beltrami conditions is that the Eqs. (\ref{be}-\ref{bh}) also serve as steady-state solutions for the equations of motion (\ref{em}), provided that the gradient terms independently become zero ($\mathbf{\nabla}\kappa _{\alpha }=0$). The subsequent condition results in $\kappa _{\alpha }=$ constant, referred to as the generalized Bernoulli condition, which indicates the homogeneity of the energy density. It is also crucial to remark that these generalized Bernoulli conditions are also consistent with Eq. (\ref{en}) \cite{Ohsaki2002}.  Therefore, the plasma equilibria characterized by the Beltrami and Bernoulli conditions are also referred to as Beltrami-Bernoulli equilibria.

 In order to obtain a relaxed state equation in terms of the magnetic or flow fields, it is necessary to solve the Ampere's law (\ref{cd}) and Beltrami conditions (\ref{be}-\ref{bh}) simultaneously. By substituting the value of $\mathbf{V}_{e}$ from Eq. (\ref{cd}) in Eq. (\ref{be}) and using Eqs. (\ref{bp}-\ref{bh}), we obtain the following value for the velocity of lighter ion species:
\begin{equation}
\mathbf{V}_{il}=\frac{1}{z_{il}N_{il}\left( b-a\right) }\left[ \left( 
\mathbf{\nabla }\times \right) ^{2}\mathbf{B}-a\mathbf{\nabla }\times 
\mathbf{B}+\gamma _{1}\mathbf{B}+z_{ih}N_{ih}\left( c-a\right) \mathbf{V}
_{ih}\right],\label{vil}
\end{equation}
where $\gamma _{1}=1+N_{p}+z_{il}^{2}M_{il}N_{il}+z_{ih}^{2}M_{ih}N_{ih}$ and $\left( \mathbf{\nabla }\times \right) ^{2}=\mathbf{\nabla }\times 
\mathbf{\nabla }\times$. By putting this value of $\mathbf{V}_{il}$ from the above relation in Eq. (\ref{bl}) and using Eq. (\ref{bh}), one can derive the following relation for $\mathbf{V}_{ih}$:
\begin{equation}
\mathbf{V}_{ih}=\frac{1}{\gamma _{3}}\left[ \left( \mathbf{\nabla }\times
\right) ^{3}\mathbf{B}-\left( a+b\right) \left( \mathbf{\nabla }\times
\right) ^{2}\mathbf{B}+\left( \gamma _{1}+ab\right) \mathbf{\nabla }\times 
\mathbf{B}-\gamma _{2}\mathbf{B}\right],\label{vih}
\end{equation}
where $\gamma _{2}=\gamma _{1}b-z_{ih}^{2}M_{ih}N_{ih}\left( c-a\right)
-z_{il}^{2}M_{il}N_{il}\left( b-a\right) $, $\gamma _{3}=z_{ih}N_{ih}\left(
b-c\right) \left( c-a\right) $ and $\left( \mathbf{\nabla }\times \right)
^{3}=\mathbf{\nabla }\times \mathbf{\nabla }\times \mathbf{\nabla }\times $. Now, by using Eq. (\ref{vih}) in Eq. (\ref{bh}), we get the following relaxed state equation in terms of magnetic field
\begin{equation}
\left( \mathbf{\nabla }\times \right) ^{4}\mathbf{B}-\zeta _{1}\left( 
\mathbf{\nabla }\times \right) ^{3}\mathbf{B}+\zeta _{2}\left( \mathbf{
\nabla }\times \right) ^{2}\mathbf{B}-\zeta _{3}\mathbf{\nabla }\times 
\mathbf{B}+\zeta _{4}\mathbf{B}=0,\label{QB}
\end{equation}
where $\zeta _{1}=a+b+c$, $\zeta _{2}=\gamma _{1}+ab+ac+bc$, $\zeta
_{3}=\gamma _{2}+c\gamma _{1}+abc$, $\zeta _{4}=c\gamma
_{2}-z_{ih}M_{ih}\gamma _{3}$ and $\left( \mathbf{\nabla }\times \right)
^{4}=\mathbf{\nabla }\times \mathbf{\nabla }\times \mathbf{\nabla }\times 
\mathbf{\nabla }\times $. The relaxed state Eq. (\ref{QB}) represents a QB field --a linear combination of four single Beltrami fields that arises from the premise of inertial behavior for all plasma species, with the same values of Beltrami parameters for pair species and distinct values for ion species. Also, the QB relaxed state in terms of flow fields can also be obtained by eliminating the magnetic field from Eqs. (\ref{cd} and \ref{be}-\ref{bh}). Significantly, the Eqs. (\ref{vil}-\ref{vih}) also indicate a strong magnetofluid coupling in the relaxed state, which is a prominent feature of the multi-Beltrami equilibrium state. An additional aspect worthy of emphasis is that, in addition to the utilization of vortex dynamics approach for the derivation of the QB field Eq. (\ref{QB}), the relaxed state equation can also be derived by the application of the variational technique. So by minimizing the following functional which includes ideal invariants for the plasma system
\begin{equation}
    \delta(E-\frac{H_{e}}{\mu_{1}}-\frac{H_{p}}{\mu_{1}}-\frac{H_{il}}{\mu_{2}}-\frac{H_{ih}}{\mu_{3}})=0,
\end{equation}
the QB state Eq. (\ref{QB}) can also be obtained, where $\mu_{1}$, $\mu_{2}$, and $\mu_{3}$ represent Lagrange's multipliers.

In order to obtain an analytical solution of the QB field Eq. (\ref{QB}), it can be expressed as a linear combination of four single Beltrami fields, i.e.,
\begin{equation}
\mathbf{B}=\sum_{j=1}^{4}C_{j }\mathbf{B}_{j},\label{sl}
\end{equation}
where $C_{j}$ are some arbitrary constants while $\mathbf{B}_{j}$ are the solutions of the following single Beltrami force-free fields
\begin{equation}
\mathbf{\nabla }\times \mathbf{B}_{j }=\lambda _{j}\mathbf{B}
_{j}.\label{sbc}
\end{equation}
Importantly, $\mathbf{B}_{j}$ can be frequently represented using either the Arnold-Beltrami-Childress (ABC) flow field in Cartesian coordinates or Chandrasekhar-Kendall eigenfunctions in cylindrical geometry \cite{Yoshida1999}. By employing Eqs. (\ref{sl}-\ref{sbc}) in Eq. (\ref{QB}), we derive the subsequent characteristic equation for a non-trivial solution of $\mathbf{B}$
\begin{equation}
\lambda ^{4}-\zeta _{1}\lambda ^{3}+\zeta _{2}\lambda ^{2}-\zeta _{3}\lambda
+\zeta _{4}=0.\label{ce}
\end{equation}
Within the framework of this plasma model, a comprehensive analysis of the quartic eigenvalue Eq. (\ref{ce}) employing discriminants, as explicated in Ref. \cite{Petrakis2008}, elucidates that merely two discernible sets of roots are possible, with the remaining root types being extremely uncommon. As a result, additional types of roots will not be addressed in this study. So Eq. (\ref{ce}), being a quartic equation, has four roots that may be either four real ($\lambda_{1}$, $\lambda_{2}$, $\lambda_{3}$, and $\lambda_{4}$)  or a combination of two real ($\lambda_{1}$ and $\lambda_{2}$) and a complex conjugate pair ($\lambda_{3}=\lambda^{*}_{4}$) in the context of this plasma system.
Additionally, the rot operator, indicated as "$\mathbf{\nabla }\times$", can be represented more succinctly as the operator "curl". By utilizing Eqs. (\ref{sl}-\ref{sbc}), the QB Eq. (\ref{QB}) can be reformulated as follows \cite{Yoshida1990}
\begin{equation}
\left( curl-\lambda _{1}\right) \left( curl-\lambda
_{2}\right) \left( curl-\lambda _{3}\right) \left(curl
-\lambda _{4}\right) \mathbf{B}=0, \label{qbe}
\end{equation}
By expanding the aforementioned equation and doing a comparison with Eq. (\ref{QB}), it can be inferred that there exists a subsequent relationship among the scale parameters and the plasma parameters, which are
\begin{subequations}
\begin{eqnarray}
\zeta _{1} &=&\lambda _{1}+\lambda _{2}+\lambda _{3}+\lambda _{4}, \label{r1}\\
\zeta _{2} &=&\lambda _{1}\left( \lambda _{2}+\lambda _{3}+\lambda
_{4}\right) +\lambda _{2}\left( \lambda _{3}+\lambda _{4}\right) +\lambda
_{3}\lambda _{4}, \\
\zeta _{3} &=&\lambda _{1}\lambda _{2}\left( \lambda _{3}+\lambda
_{4}\right) +\lambda _{2}\lambda _{3}\lambda _{4}, \\
\zeta _{4} &=&\lambda _{1}\lambda _{2}\lambda _{3}\lambda _{3}.\label{r4}
\end{eqnarray}
\end{subequations}
The relations (\ref{r1}-\ref{r4}) among $\zeta _{j=1,2,3,4}$ and $\lambda_{j=1,2,3,4}$ also adhere to Vieta's formula, so implying that the values of scale parameters equal to the roots of the quartic Eq. (\ref{ce}). The Beltrami fields, which are characterized by the eigenfunctions of the curl operator, exhibit the essential characteristics of sheared, helical, or chiral structures in different plasma settings. In the present investigation, we consider a simple slab geometry, wherein the plasma is confined inside the spatial domain $-x_{0}\leq x \geq x_{0}$ with  $\vert x \vert \leq  x_{0}> 0$. Inside this region, the magnetic field is only a function of $x$ and has only two components ($B_{y}$ and $B_{z}$). Consequently, the analytical solution of the QB state for above mentioned plasma configuration demonstrates a sheared magnetic field. So, it is also possible to express the analytical solution of Eq. (\ref{qbe}) in a simple slab geometry as
\begin{equation}
\mathbf{B}=\sum_{j=1}^{4}C_{j}[ 
sin (\lambda _{j}x)\widehat{y}+cos (\lambda _{j}x)\widehat{z}].\label{asb}
\end{equation}
In Eq. (\ref{asb}), as mentioned earlier, the constants $C_{j}$ are arbitrary and can take on either real or complex values, and their values can be determined by using the following boundary conditions: $\left\vert\mathbf{B}_{z}\right\vert_{x=0}=\sum\limits_{j=1}^{4}C_{j}=g_{1}$, $\left\vert\mathbf{B}_{y}\right\vert_{x=x_{0}}=\sum\limits_{j=1}^{4}C_{j}sin(\lambda_{j}x_{0})=g_{2}$, $\left\vert\mathbf{J}_{z}\right\vert_{x=0}=\sum\limits_{j=1}^{4}\lambda_{j}C_{j}=g_{3}$, and  $\left\vert\mathbf{J}_{y}\right\vert_{x=x_{0}}=\sum\limits_{j=1}^{4}\lambda_{j}C_{j}sin(\lambda_{j}x_{0})=g_{4}$, where $g_{1}$, $g_{1}$, $g_{1}$, and $g_{4}$ are some arbitrary and real valued constants. In Eq. (\ref{asb}), as mentioned earlier, $C_{j}$ are some arbitrary constants. In the scenario where all $\lambda_{1,2,3,4}$ are real, all of $C_{1,2,3,4}$ will be real as well. Conversely, if two real $\lambda_{1,2}$ and one pair of complex conjugate $\lambda_{3,4}$ exist, then $C_{1,2}$ come out to be real-valued, while the remaining two i.e., $C_{3,4}$ constituting a complex conjugate pair. In a three-dimensional bounded plasma domain, the Beltrami field $\mathbf{B}$ satisfies the conditions $\mathbf{n}\cdot \mathbf{B}=0$ and $\mathbf{n}\cdot \mathbf{\nabla}\times\mathbf{B}=0$, where $\mathbf{n}$ represents the unit normal vector pointing towards the smooth surface of the plasma boundary.  So the boundary conditions employed in this investigation align with the previously mentioned two conditions for a Beltrami vector field \cite{Mahajan1998,Yoshida1999}.  
\section{Results and discussion}\label{S4}
As mentioned earlier in the introduction, the presence of electron-positron multi-ion plasmas has been reported in several astrophysical settings and can also be deliberately created in laboratories. To conduct a more detailed investigation of the QB state, we will consider the plasma parameters given in Ref. \cite{Abdelwahed2020}. These parameters include the charge numbers, light positive and heavy negative ion species masses, number densities and ambient magnetic fields, which are as follows: $z_{il}=z_{ih}=1$, $m_{il}=1.67\times 10^{-24}$g, $m_{ih}=5.31\times
10^{-23}$g, $n_{e}=10^{12}$cm$^{-3}$, $n_{p}=1.4n_{e}$, $n_{il}=2n_{e}$, $n_{ih}=2.4n_{e}$ and $B_{0}=0.5-1.0$ G. Before proceeding, it is essential to note that when all scale parameters have real values, the values of $C_{1,2,3,4}$ will also be real. In this case, the analytical solution of the QB field given by Eq. (\ref{asb}) for certain boundary conditions exhibits a paramagnetic trend. The reason for this phenomenon is that the trigonometric functions will always decay away from the center. 
In contrast, the relaxed state will exhibit a diamagnetic trend when two eigenvalues of the curl operator are real and the other two are complex. Also, corresponding to this set of roots, $C_{1,2}$ will be real, while $C_{3,4}$ will be a complex conjugate pair. The fundamental reason for these diamagnetic trends is that, as a result of complex scale parameters ($\lambda_{3,4}$) and $C_{3,4}$, trigonometric functions transform into hyperbolic functions that always increase away from their origin \cite{Mahajan1998,Yoshida1999}.
\subsection{Role of generalized helicities}
As elucidated in the preceding section, the Beltrami parameters are associated with the generalized helicities of plasma species, which are regarded as the ideal invariants. In addition, they show the amplitude of the flow of plasma species, which can be Alfv\'{e}nic, sub-Alfv\'{e}nic, or super-Alfv\'{e}nic. In the present analysis, we will demonstrate the potential for the emergence of multiscale structures, as well as paramagnetic and diamagnetic structures, in the QB relaxed state through the investigation of several sets of Beltrami parameters while keeping other plasma parameters fixed ($z_{il}=z_{ih}=1$, $m_{il}=1.67\times 10^{-24}$g, $m_{ih}=5.31\times
10^{-23}$g, $n_{p}=1.4n_{e}$, $n_{il}=2n_{e}$, $n_{ih}=2.4n_{e}$).
\begin{itemize}
\item Consider the scenario in which the flows of all plasma species are Alfv\'{e}nic ($a\approx b \approx c \approx 1$) and the values of the Beltrami parameters are $a=1.0$, $b=1.1$ and $c=1.15$. In this particular case, the eigenvalues are $\lambda_{1}=1.097$, $\lambda_{2}=1.15$ and $\lambda_{3,4}=0.5012\pm 1.4803i$, respectively. The presence of both real and complex eigenvalues gives rise to the emergence of diamagnetic structures. Additionally, it is important to acknowledge that the dimensions of the self-organized vortices are comparable to the characteristic length scale $l_{e}$, i.e., $\left\vert\lambda_{1,2}\right\vert^{-1}\approx l_{e}$ and $\left\vert\lambda_{3,4}\right\vert^{-1}= 1.5l_{e}$. Within the context of plasma self-organization, microscale structures are defined as relaxed state structures on the order of electron skin depth. In contrast, macroscale structures are those that are significantly larger than the characteristic length scale. Therefore, in the case described above, all the relaxed structures are microscale structures. Importantly, such microscale relaxed state structures play a crucial role in the time-dependent dynamo and reverse-dynamo processes, featuring the generation of large scale magnetic fields and fast flows \cite{Mahajan2005,Lingam2015}.
\item In situations where the plasma species have sub-Alfv\'{e}nic flows ($a\approx b \approx c \gg 1$), the scale parameters for the given Beltrami parameters ($a=20.0$, $b=20.5$ and $c=20.9$) are determined to be $\lambda_{1}=0.13$, $\lambda_{2}\approx a$, $\lambda_{3}\approx b$ and $\lambda_{4}\approx c$. Since all the scale parameters possess real values, the QB state will have a paramagnetic behavior. Furthermore, based on the scale parameters, it is obvious that one of the self-organized structures has a significantly bigger size compared to $l_{e}$ ($\left\vert\lambda_{1}\right\vert^{-1}\gg l_{e}$ ), but the other three structures are considerably smaller than $l_{e}$ ($\left\vert\lambda_{2,3,4}\right\vert^{-1}\ll l_{e}$). 
\item When the flows of plasma species are super-Alfv\'{e}nic ($a\approx b \approx c \ll 1$), i.e., $a=0.1$, $b=0.15$ and $c=0.19$, two scale parameters exhibit real values ($\lambda_{1}\approx b$ and $\lambda_{2}\approx c$) while the remaining two are complex ($\lambda_{3,4}=0.0507\pm 1.5621i$). In accordance with the aforementioned scale parameters, it can be observed that the dimensions of two relaxed state structures exceed that of $l_{e}$, but the dimensions of the remaining two structures are on the order of $l_{e}$ ($\left\vert\lambda_{1,2}\right\vert^{-1}\gg l_{e}$ and $\left\vert\lambda_{3,4}\right\vert^{-1}\approx l_{e}$). Furthermore, it can be observed that the relaxed state shows a diamagnetic behavior in the context of super-Alfv\'{e}nic plasma flows.
\item When $a=0.1$, $b=20.1$ and $c=20.5$, which means that the flows of pair species are super-Alfv\'{e}nic ($a\ll 1$) and the flows of ion species are sub-Alfv\'{e}nic ($ b \approx c \gg 1$), then two scale parameters are real and the other two are real with following values: $\lambda_{1}\approx b$, $\lambda_{2}\approx c$ and $\lambda_{3,4}=0.051\pm 1.5482i$. As a result of these eigenvalues, the relaxed state will exhibit a diamagnetic characteristic. In this particular situation, the scale hierarchy appears as follows: $\left\vert\lambda_{1,2}\right\vert^{-1}\ll l_{e}$ and $\left\vert\lambda_{3,4}\right\vert^{-1}\approx l_{e}$.
\item When the values of $a$, $b$ and $c$ given as $a = 10.0$, $b = 0.1$ and $c = 0.15$, respectively, it can be inferred that the flows of pair species are sub-Alfv\'{e}nic ($a\gg 1$), while the flows of ion species are super-Alfv\'{e}nic ( $b\approx c \ll 1$). Under these conditions, it is observed that two scale parameters are real, while the other two are complex conjugate, and the values of scale parameters are $\lambda_{1}\approx a$, $\lambda_{2}\approx b$ and $\lambda_{3,4}=0.1729\pm 0.196i$. The diamagnetic feature of the relaxed state will be observed due to the presence of these eigenvalues. The scale hierarchy in this circumstance is as follows:$\left\vert\lambda_{1}\right\vert^{-1}\ll l_{e}$, $\left\vert\lambda_{2}\right\vert^{-1}\gg l_{e}$ and $\left\vert\lambda_{3,4}\right\vert^{-1}\gtrsim l_{e}$.
\end{itemize}
Based on the above discussion, it is evident that the variation of Beltrami parameters can induce a transition in the nature of relaxed state structures. Additionally, the variation in the values of Beltrami enables the creation of multiscale structures. The presence of multiscale structures offers the potential for field and flow variations over different length scales, which can result in the heating of plasma \cite{Mahajan2001}. Moreover, the presence of microscale structures within the QB state has the potential to serve as an energy source for the creation of large scale fields and flows \cite{Mahajan2005}.
\subsection{Role of plasma species densities}
The impact of plasma species density is also significant in the creation of relaxed-state structures. In order to demonstrate the influence of plasma species density on the nature of self-organization, we keep constant values for the Beltrami parameters ($a=5.0$, $b=0.9$ and $c=1.0$) while varying the densities of the plasma species. As an illustration, in the scenario where $N_{p}=1.4$, $N_{il}=2.0$ and $N_{ih}=2.4$, the corresponding values of the scale parameters are $\lambda_{1}=0.589$, $\lambda_{2}=0.7483$ and $\lambda_{3,4}=0.051\pm 1.5482i$ .  The presence of both real and complex eigenvalues allows for the formation of diamagnetic structures within the QB state. Conversely, in the scenario where the density of positron species is decreased, i.e., $N_{p}=1.0$, $N_{il}=2.4$ and $N_{ih}=2.4$, the resulting scale parameters are as follows: $\lambda_{1}=0.589$, $\lambda_{2}=0.7483$, $\lambda_{3}=0.9999$ and $\lambda_{4}=4.56$. Due to the fact that all scale parameters possess real values, the relaxed state will exhibit paramagnetic features. In a similar vein, increasing the concentration of heavy negative ion species can lead to transformations in the characteristics of scale parameters. As an example, when the values of $N_{p}$, $N_{il}$ and $N_{ih}$ are $1.4$, $2.5$ and $2.9$ respectively, the scale parameters are found to be $\lambda_{1}=0.9999$, $\lambda_{2}=4.4637$ and $\lambda_{3,4}=0.7181\pm 0.1587i$. In accordance with these eigenvalues, relaxed state structures will demonstrate diamagnetic trends. Based on the aforementioned analysis, it can be inferred that by increasing the densities of positrons and heavy negative ion species, the relaxed state's paramagnetic characteristics can be transformed into diamagnetic ones. Furthermore, these transformations (paramagnetic into diamagnetic and vice versa) hold significance within the framework of energy conversion mechanisms in space plasmas.
\subsection{Role of inertia of pair and ion species}
The inertial effects of plasma species also have a significant role in the self-organized states of this plasma system. For instance, when the inertia of pair species (electrons and positrons) is ignored, one can derive a triple Beltrami (TB) relaxed state equation that can be expressed as
\begin{equation}
\left( \mathbf{\nabla }\times \right) ^{3}\mathbf{B}-\zeta _{1}\left( 
\mathbf{\nabla }\times \right) ^{2}\mathbf{B}+\zeta _{2}\mathbf{
\nabla }\times\mathbf{B}-\zeta _{3} 
\mathbf{B}=0,\label{TB1}
\end{equation}
where $\zeta _{1}=((n_{p}-n_{e})/an_{il})+b+c$, $\zeta _{2}=1+(m_{il}n_{ih}/m_{ih}n_{il})+b(m_{il}n_{ih}/m_{ih}n_{il})+(c(n_{p}-n_{e})/an_{il})+bc$ and $\zeta _{3}=c+b(b-c)(m_{il}n_{ih}/m_{ih}n_{il})+bc((n_{p}-n_{e})/an_{il})$. It is crucial to note that the TB self-organized state can be expressed as the linear combination of three different single Beltrami states. As a result, this state has three distinct eigenvalues that consequently enable the formation of three self-organized vortices. Furthermore, the eigenvalues can exhibit either real values or a single real eigenvalue, with the remaining two eigenvalues forming complex conjugate pairs. The paramagnetic behavior of the TB relaxed state is exhibited when all three eigenvalues are real, but a diamagnetic trend can be seen when there is a pair of complex conjugate eigenvalues along with a real one. Also, the presence of three different eigenvalues in the TB state enables the emergence of multiscale structures, in addition to the paramagnetic and diamagnetic behaviors.

On the other hand, if the generalized helicities of ion species are also assumed to be equal, the relaxed state turns out to be a double Beltrami (DB) state, which is a linear superposition of two linear force-free Beltrami fields with two relaxed state structures. The DB state is characterized by two eigenvalues that may take on the values of real, complex, or imaginary numbers. Paramagnetic structures are formed when the eigenvalues are real, but the presence of complex eigenvalues gives rise to either a partial diamagnetic structure or a diamagnetic structure \cite{Mahajan1998}. In contrast, it is possible to build perfect diamagnetic structures by considering pure imaginary eigenvalues \cite{Mahajan2008}. In addition, there will be a loss of equilibrium when both of the real eigenvalues are equal, which will result in the catastrophic transformation of magnetic energy into kinetic energy \cite{Ohsaki2002}. So in the context of the plasma system being discussed, and considering the conditions mentioned earlier regarding generalized helicities, the DB relaxed state equation can be formulated as follows
\begin{equation}
\left( \mathbf{\nabla }\times \right) ^{2}\mathbf{B}-\zeta _{1} 
\mathbf{\nabla }\times\mathbf{B}+\zeta _{2}\mathbf{B}=0.\label{DB1}
\end{equation}
where $\zeta _{1}=((n_{p}-n_{e})/an_{il})+b$ and $\zeta _{2}=1+(m_{il}n_{ih}/m_{ih}n_{il})+b((n_{p}-n_{e})/an_{il})$. Also, in Eqs. (\ref{TB1}-\ref{DB1}) length and plasma species flows are normalized with lighter ion skin depth and Alfv\'{e}n speed. In contrast to the preceding case, when we consider ion species as static and inertial pair species having the same generalized helicities, the resulting self-organized state is referred to as the DB state, which can be represented by the following equation
\begin{equation}
\left( \mathbf{\nabla }\times \right) ^{2}\mathbf{B}-a 
\mathbf{\nabla }\times\mathbf{B}+(1+N_{p})\mathbf{B}=0.\label{DB2}
\end{equation}
However, if we consider the distinct generalized helicities of pair species, one can derive the following TB state equation
\begin{equation}
\left( \mathbf{\nabla }\times \right) ^{3}\mathbf{B}-\zeta _{1}\left( 
\mathbf{\nabla }\times \right) ^{2}\mathbf{B}+\zeta _{2}\mathbf{
\nabla }\times\mathbf{B}-\zeta _{3} 
\mathbf{B}=0,\label{TB2}
\end{equation}
where $\zeta _{1}=a+b$, $\zeta _{2}=1+N_{p}+ab$ and $\zeta _{3}=b+aN_{p}$. It is abundantly clear from the preceding discussion that the inertia of plasma species plays a crucial role in the relaxation process. Lower-index self-organized states (DB and TB) can also be derived for this plasma model by ignoring the inertia of either pair species or ion species.
\section{Summary and Conclusion}\label{S5}
The present study focuses on the investigation of a NFF QB relaxed state of a magnetized multispecies plasma, consisting of mobile electrons, positrons, light positive ions and heavy negative ions. In deriving the QB field equation, we considered the same generalized helicities for pair species, but distinct generalized helicities for ion species. The QB state is characterized by four scale parameters that can be either real or a combination of both real and complex conjugate pair, that enables the construction of four multiscale self-organized vortices.
In addition, these scale parameters also determine the nature of the relaxed state, i.e., for the real scale parameters, the QB state exhibits a paramagnetic trend, whereas for the combination of real and complex values, it exhibits a diamagnetic trend. The investigation of the QB state with some model plasma parameters typical of the earth's ionosphere reveals that multiscale structures can be formed by adjusting the generalized helicities of plasma species for given number densities. Similarly, the manipulation of positron and heavy negative ion species densities, while keeping the generalized helicities of plasma species fixed, enables the transformation of paramagnetic trends into diamagnetic trends, and vice versa. The densities of plasma species exhibit variations that are associated with cosmic-ray and solar flare activity. Consequently, the changes in density can facilitate the conversion of paramagnetic structures into diamagnetic ones, and vice versa. This phenomenon holds potential for enhancing our comprehension of ionospheric plasma dynamics. Furthermore, it has been demonstrated that the inertia of both pair and ion species significantly affects the self-organized state of the plasma. 
Particularly, if we neglect the inertia of pair species, the resulting relaxed state is a TB state. However, if we consider the pair species to have inertia with distinct generalized helicities and assume the ion species to be static, the relaxed state is a TB state. Conversely, for the same helicities of pair species, the relaxed state becomes a DB state. The present study will contribute to a deeper comprehension of space plasmas, particularly the ionosphere of the earth, alongside laboratory-based plasma investigations encompassing multispecies plasmas featuring negative ions.


\begin{thebibliography}{999}

\bibitem{Ortolani} S. Ortolani and D. D. Schnack, Magnetohydrodynamics of Plasma Relaxation (World Scientific, Singapore, 1993).

\bibitem{Woltjer1958} L. Woltjer, Proc. Natl. Acad. Sci. 44, 489 (1958).

\bibitem{Taylor1974} J. B. Taylor, Phys. Rev. Lett. 33, 1139 (1974).

\bibitem{Steinhauer1997} L. C. Steinhauer and A. Ishida, Phys. Rev. Lett. 79, 3423 (1997).

\bibitem{Mahajan1998} S. M. Mahajan and Z. Yoshida, Phys. Rev. Lett. 81, 4863 (1998).

\bibitem{Yoshida1999} Z. Yoshida and S.M. Mahajan, J. Math. Phys. 40, 5080 (1999).

\bibitem{Steinhauer1998} L. C. Steinhauer and A. Ishida, Phys. Plasmas 5, 2609 (1998).

\bibitem{Steinhauer2002} L. C. Steinhauer, Phys. Plasmas 9, 3767 (2002).

\bibitem{Yoshida2002} Z. Yoshida and S. M. Mahajan, Phys. Rev. Lett. 88, 095001 (2002).

\bibitem{Mahajan2015} S. M. Mahajan and M. Lingam, Phys. Plasmas 22, 092123 (2015).

\bibitem{Mahajan2000} S. M. Mahajan and Z. Yoshida, Phys. Plasmas 7, 635 (2000).

\bibitem{Yoshida2001} Z. Yoshida, S. M. Mahajan, S. Ohsaki, M. Iqbal, and N. Shatashvili, Phys. Plasmas 8, 2125 (2001).

\bibitem{Mahajan2001} S. M. Mahajan, R. Miklaszewski, K. I. Nikol’skaya, and N. L. Shatashvili, Phys. Plasmas 8, 1340 (2001).

\bibitem{Mahajan2002} S. M. Mahajan, K. I. Nikol’skaya, N. L. Shatashvili, and Z. Yoshida, Astrophys. J. 576, L161 (2002).

\bibitem{Ohsaki2002} S. Ohsaki, N. L. Shatashvili, Z. Yoshida, and S. M. Mahajan, Astrophys. J. 570, 395 (2002).

\bibitem{Bhattacharyya2007} R. Bhattacharyya, M. S. Janaki, B. Dasgupta, and G. P. Zank, Sol. Phys. 240, 63 (2007).

\bibitem{Kumar2011} D. Kumar and R. Bhattacharyya, Phys. Plasmas 18, (2011).

\bibitem{Mahajan2005} S. M. Mahajan, N. L. Shatashvili, S. V. Mikeladze, and K. I. Sigua, Astrophys. J. 634, 419 (2005).

\bibitem{Lingam2015} M. Lingam and S. M. Mahajan, Mon. Not. R. Astron. Soc. Lett. 449, L36 (2015).

\bibitem{Abdelhamid2016a} H. M. Abdelhamid and Z. Yoshida, Phys. Plasmas 23, 022105 (2016).

\bibitem{Abdelhamid2016b} H. M. Abdelhamid, M. Lingam, and S. M. Mahajan, Astrophys. J. 829, 87 (2016).

\bibitem{Abdelhamid2017} H. M. Abdelhamid and Z. Yoshida, Phys. Plasmas 24, 022107 (2017).

\bibitem{Mahajan2020} S. M. Mahajan and M. Lingam, Mon. Not. R. Astron. Soc. 495, 2771 (2020).

\bibitem{Berezhiani2015} V. I. Berezhiani, N. L. Shatashvili, and S. M. Mahajan, Phys. Plasmas 22, 022902 (2015).

\bibitem{Shatashvili2016} N. L. Shatashvili, S. M. Mahajan, and V. I. Berezhiani, Astrophys. Space Sci. 361, 70 (2016).

\bibitem{Shatashvili2019} N. L. Shatashvili, S. M. Mahajan, and V. I. Berezhiani, Astrophys. Space Sci. 364, 148 (2019).

\bibitem{Usman2021} U. Shazad, M. Iqbal, and S. Ullah, Phys. Scr. 96, 125627 (2021).

\bibitem{Usman2023a} U. Shazad and M. Iqbal, Phys. Scr. 98, 055605 (2023).

\bibitem{Usman2023b} U. Shazad and M. Iqbal, Z. Naturforsch. A 78, 983 (2023).

\bibitem{Usman2024} U. Shazad and M. Iqbal, Braz. J. Phys. 54, 22 (2024).

\bibitem{Bhattacharjee2018} C. Bhattacharjee, J. C. Feng, and D. J. Stark, Mon. Not. R. Astron. Soc. 481, 206 (2018).

\bibitem{Asenjo2019} F. A. Asenjo and S. M. Mahajan, Phys. Rev. E 99, 053204 (2019).

\bibitem{Bhattacharjee2020a} C. Bhattacharjee and J. C. Feng, Phys. Plasmas 27, 072901 (2020).

\bibitem{Bhattacharjee2020b} C. Bhattacharjee, Phys. Lett. A 384, 126698 (2020).

\bibitem{Shafa2022} S. Ullah, U. Shazad, and M. Iqbal, Phys. Scr. 97, 065605 (2022).

\bibitem{Faheem2023} F. Ahmed, M. Iqbal, and U. Shazad, AIP Adv. 13, 055305 (2023).

\bibitem{Bhattacharjee2023} C. Bhattacharjee, Phys. Rev. E 107, 035207 (2023).

\bibitem{Usman2023c} U. Shazad and M. Iqbal, J. Plasma Phys. 89, 905890512 (2023).

\bibitem{Vladimirov2003} S. V. Vladimirov, K. Ostrikov, M. Y. Yu, and G. E. Morfill, Phys. Rev. E 67, 036406 (2003).

\bibitem{Adriani2009} O. Adriani, G. C. Barbarino, and G. A. Bazilevskaya, Nature 458, 607 (2009).

\bibitem{Kourakis2006} I. Kourakis, A. Esfandyari-Khalejahi, M. Mehdipoor, and P. K. Shukla, Phys. Plasmas 13, 052117 (2006).

\bibitem{Massey1976} H. Massey, Negative Ions, 3rd ed. (Cambridge University Press, Cambridge,
1976).

\bibitem{Chaizy1991} P. Chaizy, H. Rème, J. A. Sauvaud, C. d’Uston, R. P. Lin, D. E. Larson, D. L. Mitchell, K. A. Anderson, C. W. Carlson, A. Korth, and D. A. Mendis, Nature 349, 393 (1991).

\bibitem{Coates2007} A. J. Coates, F. J. Crary, G. R. Lewis, D. T. Young, J. H. Waite, Jr., and E. C. Sittler, Jr., Geophys. Res. Lett. 34, L22103 (2007).

\bibitem{Ichiki2002} R. Ichiki, S. Yoshimura, T. Watanabe, Y. Nakamura, and Y. Kawai, Phys. Plasmas 9, 4481 (2002).

\bibitem{Bacal1979} M. Bacal and G. W. Hamilton, Phys. Rev. Lett. 42, 1538 (1979).

\bibitem{Sheehan1988} D. P. Sheehan and N. Rynn, Rev. Sci. Instrum. 59, 1369 (1988).

\bibitem{Gottscho1986} R. A. Gottscho and C. E. Gaebe, IEEE Trans. Plasma Sci. 14, 92 (1986).

\bibitem{Sultana2014} S. Sultana and A. A. Mamun, Astrophys. Space Sci. 349, 229 (2014).

\bibitem{Jannat2015} N. Jannat, M. Ferdousi, and A. A. Mamun, J. Korean Phys. Soc. 67, 496 (2015).

\bibitem{Jannat2016} N. Jannat, M. Ferdousi, and A. A. Mamun, Plasma Phys. Reports 42, 678 (2016).

\bibitem{Chowdhury2017} N. A. Chowdhury, A. Mannan, M. M. Hasan, and A. A. Mamun, Chaos 27, 093105 (2017).

\bibitem{Ahmed2018} N. Ahmed, A. Mannan, N. A. Chowdhury, and A. A. Mamun, Chaos 28, 123107 (2018).

\bibitem{Khondaker2019} S. Khondaker, A. Mannan, N. A. Chowdhury, and A. A. Mamun, Contrib. to Plasma Phys. 59, 1-9 (2019).

\bibitem{Abdelwahed2020} H. G. Abdelwahed, R. Sabry, and A. A. El-Rahman, Adv. Sp. Res. 66, 259 (2020).

\bibitem{Douanla2020} D. V. Douanla, D. V. Alim, C. G. L. Tiofack, and A. Mohamadou, Contrib. to Plasma Phys. 60, 1-14 (2020).

\bibitem{Jahan2021} S. Jahan, M. N. Haque, N. A. Chowdhury, A. Mannan, and A. Al Mamun, Universe 7, 63 (2021).

\bibitem{Taibany2021} W. F. El-Taibany, N. A. El-Bedwehy, N. A. El-Shafeay, and S. K. El-Labany, Galaxies 9, 48 (2021).

\bibitem{Heera2021} N. M. Heera, J. Akter, N. K. Tamanna, N. A. Chowdhury, T. I. Rajib, S. Sultana, and A. A. Mamun, AIP Adv. 11, 055117 (2021).

\bibitem{Tajima1990} T. Tajima and T. Taniuti, Phys. Rev. A 42, 3587 (1990).

\bibitem{Petrakis2008} A.L. Petrakis and L.A. Petrakis, J. Interdiscip. Math. 11, 815 (2008).

\bibitem{Yoshida1990} Z. Yoshida and Y. Giga, Math. Zeitschrift 204, 235 (1990).

\bibitem{Mahajan2008} S.M. Mahajan, Phys. Rev. Lett. 100, 075001 (2008).

\end{thebibliography}
\end{document}